\documentstyle[12pt,a4,epsfig]{elsarticle}

\newlength{\abstwidth}
\begin{document}
\thispagestyle{empty}
\begin{center}
{\Large {\bf Correlations and Event-by-Event Fluctuations in High Multiplicity Events Produced in $^{208}$Pb-$^{208}$Pb Collisions }}\\[5mm]
\end{center}
{\bf Shakeel Ahmad$^a$\footnote{email: Shakeel.Ahmad@cern.ch\\ {\bf Authors declare that there is no conflict of interest}}, Shaista Khan$^a$, Ashwini Kumar$^b$, Arpit Singh$^c$, A. Ahmad$^d$ and B. K. Singh$^c$}\\
\begin{enumerate}
\item[a)]{\it  Department of Physics, Aligarh Muslim University, Aligarh-202002, INDIA}\\[-5mm]
\item[b)] {\it Department of Physics and Electronics, Dr. Ram Manohar Lohia Avadh University, Faizabad - 224001, INDIA} \\[-2mm]
\item[c)]{\it Department of Physics, Banaras Hindu University, Varanasi - 221005, INDIA} \\[-2mm] 
\item[d)] {\it  Department of Applied Physics, Aligarh Muslim University, Aligarh-202002, INDIA}\\[-2mm]
\end{enumerate}

\begin{center}
\newpage
{\bf Abstract}\\[2ex]
\begin{minipage}{\abstwidth} Analysis of high multiplicity events produced in 158A GeV/c $^{208}$Pb-$^{208}$Pb collisions is carried out to study the event-by-event fluctuations. The findings reveal that the method of scaled factorial moments can be used to identify the events having densely populated narrow phase space bins. A few events sorted out adopting this approach are individually analyzed. It is observed that these events do exhibit large fluctuations in their pseudorapidity, $\eta$ and azimuthal angle, $\phi$ distributions arising out due to some dynamical reasons. Two particle $\Delta\eta$-$\Delta\phi$ correlation study applied to these events too indicates that some complex two-dimensional structure of significantly high magnitude are present in these events which might have some dynamical origin. The findings reveal that the method of scaled factorial moments may be used as an effective triggering for events with large dynamical fluctuations. \\

\end{minipage}
\end{center}

\noindent KEY-WORDS: Event-by-event fluctuations, Correlations, Multiparticle production, Relativistic heavy-ion collisions.\\
\phantom{dummy}

\newpage
\noindent {\bf 1. Introduction:}\\
\noindent Investigations involving fluctuations in collisions of heavy nuclei at relativistic energies might serve as a useful tool to identify the existence of the state of partonic matter in early life of the fireball because of the idea that the fluctuations in a thermal system are directly related to various susceptibilities and would be a good indication for the possible phase changes[1,2,3,4]. Furthermore, large event-by-event (ebe) fluctuations of the suitably chosen observables in ion-ion (AA) collisions would help identifying events of distinct classes, e.g. one with and another without quark-gluon plasma (QGP)[2], as under extreme conditions of temperature and energy density,  a novel phase of matter--the QGP is expected to be produced. The search for the occurrence of phase transition from hadronic matter to QGP still remains a favorite topic for high energy physicists[1,5,6]. It is commonly believed that even if the extreme conditions of QGP formation are achieved in relativistic AA collisions, not all the events would be produced via QGP as the cross-section for the QGP formation[1] is still not known. Therefore, a chosen sub-sample of events exhibiting large fluctuations in certain observables should be studied in detail[7,8]. A significant contribution to the observed fluctuations in a variable comes due to finite number of particles produced in an event and is referred to as the statistical fluctuations. Its magnitude could be evaluated by considering the independent emission of particles or by using the event mixing technique[1,2,7,8]. All other fluctuations are of dynamical origin and are divided into two groups: a) fluctuations which do not change on ebe basis, e.g. two-particle correlation due to Bose-Einstein statistics or due to decay of resonances and b) fluctuations which change on ebe basis and are termed as ebe fluctuations. Examples are fluctuations in charged to neutral particle multiplicity ratio due to creation of disoriented chiral condensate (DCC) region or creation of jets which contribute to the high p$_{t}$ tail of transverse momentum distributions[1,2]. Several attempts[1,2,3,9,10,11,12,13,14] have been made to investigate ebe fluctuations in heavy-ion collisions at relativistic energies. The findings do indicate the presence of dynamical fluctuations.\\
\noindent The main aim of the present study is to search for the rare events exhibiting some unusual behaviour from a data sample with large number of events. The analyses of individual collision events at SPS and RHIC energies have been argued to statistically reliable as the multiplicities of these events would be high enough and the statistical fluctuations may be treated as under control[15,16,17,18,19,20,21,22]. Moreover, from the study of such events with strong fluctuations, information about the dominance of dynamical components of fluctuations over the statistical ones can be extracted, which, in turn, would provide more insight into the underlying dynamics of high multiplicity events[23,24]. Analysis of one "hadron-rich" event[25] in the context of Centauro event, single event p$_{t}$ distributions[26], single event k/$\pi$ ratio[27], intermittency in individual events, entropy and cluster analysis in single events[21,22], etc., are some of the investigations carried out so far.\\

 An attempt, therefore, has been made to carry out the analysis of a few high multiplicity individual events produced in 158A GeV/c Pb-Pb collisions. These events are taken from the emulsion experiment performed by EMU01 Collaboration[28,29,30,31].It should be emphasized that the conventional emulsion technique has an advantage over the other detectors due to its 4$\pi$ solid angle coverage and the data are free from biases due to full phase space acceptance whereas other detectors have limited acceptance cone. This not only reduces the charged particle multiplicity but might also distort some of the event characteristics, like particle density fluctuations. Furthermore, to test whether the fluctuations are arising due to non-statistical reasons, the findings are compared with the reference distributions obtained by event mixing technique[15,16,32].\\
\noindent {\bf 2. Results and Discussion:}\\
\noindent {\bf 2.1 Single event factorial moments}\\
\noindent Method of scaled factorial moments (SFMs)[33,34,35], has been extensively used [1,15,36] to search for the non-linear phenomena in hadronic and ion-ion collisions at widely different energies. The power law behaviour of the type $F_{q} \sim M^{\psi}_{q}$ has been investigated by studying[37,38] the horizontally averaged vertical moments F$_{q}^{v}$ and (or) the vertically averaged horizontal moments, F$_{q}^{h}$. The results suggest the presence of large particle density fluctuations in narrow phase space bins to be rare that can not be ignored. It has, however, been pointed out[37,39] that because of the averaging procedure adopted, studies involving F$_{q}^{v}$ or F$_{q}^{h}$ may not fully account for all the fluctuations a system might exhibit. Moreover, the values of the factorial moments F$_{q}^{e}$, estimated on ebe basis has been observed to exhibit large fluctuations and hence a distinct distribution of F$_{q}^{e}$ for a given order q and bin width, $\delta$ may be obtained for a given data sample[18,37,39]. This result suggests that the method of SFMs may be applied to individual events[15,40,41] to search for the non-linear phenomena in 'hot' and 'cold' events; the 'hot' and 'cold' events refer to the events respectively with and without large particle densities in narrow phase space bins. Method of SFMs applied to individual high multiplicity cosmic ray events[41] does indicate the presence of dynamical fluctuations. Such analysis, if carried out on ebe basis would help selecting the events with large dynamical fluctuations, if this property is not a typical one for each single event. These suppositions have, therefore, been tested in the present study by applying the SFMs analysis to the individual events.\\
\noindent The event factorial moments of order q is defined as
\begin{eqnarray}
F_q^{e} = \frac{<n(n-1)....(n-q+1)>_e}{<n>_e^q}
\end{eqnarray}
\noindent where n denotes the multiplicity in a particular $\eta$ or $\phi$ bin.\\

\noindent Values of F$_{q}^{e}$ for q =2 are calculated for each of 47 events considered by varying the number of cells, M. Since the number of events are limited, instead of plotting the F$_{2}^{e}$ distributions; we have plotted event wise \(lnF_{2}\) values for M = 10 and 30 (in the pseudo-rapidity ($\eta$ = -ln{tan$\theta$/2}) and the azimuthal angle ($\phi$) spaces); $\theta$ being the emission angle of a charged particle with respect to the mean beam direction. For the values, M = 25 and 100, we have shown the plot in two-dimensional $\eta$-$\phi$ space. These plots are shown in Fig.1. Values of \(lnF_{2}\) for the corresponding mixed events are also displayed in the same figure for comparison sake. It may be noted from the figure that \(lnF_{2}\) values of mixed events are nearly the same $\sim$ 0.3 whereas these values for the real data vary from event to event such that one would get a distinct distribution of \(lnF_{2}\) for a significant number of events.  It is quite interesting to notice that the values of \(lnF_{2}\) for few of the real events are much higher as compared to the average values taken over all the events or the values from the mixed events. These observations, thus, help identifying a few events exhibiting significantly large F$_{2}$ values in $\eta$-, $\phi$- or $\eta$-$\phi$ space for further analysis. Four such events picked-up are labeled as Evt$\#$ 5, 16, 21 and 27. These events henceforth, would be termed as 'hot' events. The multiplicities of these events are respectively 932, 852, 1433 and 974. Yet another event, Evt$\#$22 having F$_{2}$ values close to the mixed events values has also been taken for comparison sake. This event henceforth will be referred to as the 'cold' event. The multiplicity of this event is 1480. Variations of \(lnF_{2}\) with \(lnM\) for these five real and corresponding mixed events in $\eta$-, $\phi$- or $\eta$-$\phi$ spaces are displayed in Fig.2. Event averaged values, \(<lnF_{2}>\) against \(lnM\) for the entire event sample are also plotted in the same figure. It is evidently clear from Fig.2 that F$_{2}$ values for the four 'hot' events are significantly larger than that of the 'cold' event or the event averaged values. These findings, therefore, tend to suggest a few events with high density phase region or strong fluctuations are present in the real data and that these fluctuations might have some dynamical origin. These dynamical fluctuations are the reflection of the dynamics and response of the system and are indicative of phase transition like phenomenon occurring under extreme conditions of heavy ion collision such as the conditions achieved here at 158 A GeV/c.  This result, in turn, leads us to conclude that the method of SFMs, if applied to the single event, may be useful for selecting the preliminary events where strong dynamical fluctuations might be present and thereafter more advanced triggering might occur, e.g. particle ratios, enhanced particle multiplicities in certain kinematical regions, etc., may be applied.\\

\noindent {\bf 2.2 $\eta$ and $\phi$ distribution of single event}\\
\noindent By examining the F$_{2}$ values on ebe basis it might be possible to identify the rare events having high density phase space regions. Therefore, to check further that the four 'hot' events identified as above on the basis of their F$_{2}$ values do have the densely populated phase space regions, $\eta$- and $\phi$ distributions of these events along with the 'cold' event are plotted in Fig.3, whereas two-dimensional $\eta$-$\phi$ distributions for these events are exhibited in Fig.4 and Fig.5. Moreover in order to ensure that the observed spikes and valleys are the event characteristics and are not due to statistical reasons $\eta$- and $\phi$- distributions of the corresponding five mixed events are also plotted in the same figures. It is interesting to note in these figures that one-dimensional $\eta$ and $\phi$ distributions as well as two-dimensional $\eta$-$\phi$ distributions of the four 'hot' events do exhibit distinct peaks and valleys while the fifth event (the 'cold' event) show no such spikes and the distributions of these events match with those obtained from the corresponding mixed events. It may also be noted from the figure that the particle densities in the spiky regions are larger than the expected average values by a factor of $\sim$2.\\

\noindent {\bf 2.3 Two Particle Correlations}\\
\noindent Studies involving multiparticle correlations are widely accepted as a tool to search for the occurrence of phase transition in relativistic AA collisions[42,43]. As the presence of the correlations among emitted particles guides us towards the fluctuation occurred in the observables during the phase transition. It has been reported[43,44] that the inclusive two particle correlations have two components: the direct two particle correlations conventionally referred to as the short-range correlations (SRC) and the effective long-range correlations (LRC). The strong SRC have been reported to be present in several investigations[42,43,44]. These correlations have been reported to remain confined to a region, $\eta$ $\pm$ 1 units around mid-rapidity. Their properties have been explored by the concept of clustering[44]. LRC, on the other hand, arise due to ebe fluctuations of overall multiplicity and are expected[43,44] to extend over a relatively longer range ($>$2 units of $\eta$). The idea of particle production through the formation of clusters, rather than the independent emission, has been widely adopted in describing the various features of particle production[44,45]. It is widely accepted that the SRC arise due to the tendency of hadrons to be grouped in clusters, which are formed during the intermediate stage of the collision. The cluster formed are independently emitted according to a dynamically generated distribution in $\eta$ and $\phi$ and then decay isotropically in their own rest frame into the real physical particles[43,44] finally measured in the detectors. The observed two-particle correlations would permit to disentangle different correlation sources which can be directly connected to the phenomena like, collective flow, jets, resonance decays, etc.[46].\\

\noindent Two particle angular correlations in $\eta$ and $\phi$ spaces were first studied by ACM Collaboration at ISR energies[47]. It was observed that the dominant contribution to the correlation comes from the two- and three-body decay of resonances ($\eta$, $\rho^{0}$, $\omega$). Two structures were discovered, i) an enhancement near $\Delta\phi = \pi$ (away-side) explained by the two-body decay scenario and ii) the enhancement at $\Delta\phi \simeq 0$ together with an azimuthal ridge (centered at $\Delta\eta = 0$) consistent with three-body decays. These features were, later on, confirmed by PHOBOS Collaboration[44]. Fluctuations observed in the physically measurable quantities provide us indirect measure of the width of the two particle densities. Thus, these are quite useful in providing us additional informations in comparison to those obtained by studying their averages.    Besides this a significant contribution to the two particle correlations comes from the collective effects. They appear as a modulation in $\Delta\phi$ and are usually searched for in high multiplicity events[46,48]. An attempt is, therefore, made to study the two particle correlations in individual events exhibiting large fluctuations in their $\eta$ and $\phi$ distributions which have been sorted out as discussed in the previous sections. The findings are compared with the corresponding mixed events which would help extracting the contributions present due to the dynamical reasons.\\

\noindent Two particle correlations are generally studied in terms of the differences in $\eta$ and $\phi$ values between the two particles produced in the same event[46].\\
\begin{eqnarray}
\Delta\eta = |\eta_{1} - \eta_{2}|, \hspace{3ex} \Delta\phi = |\phi_{1} - \phi_{2}|
\end{eqnarray}

\noindent The correlation function C($\Delta\eta$,$\Delta\phi$) is defined and calculated as:
\begin{eqnarray}
C(\Delta\eta,\Delta\phi) = \frac{N^{pair}_{mixed}  \rho^{II}_{d}(\Delta\eta,\Delta\phi)}{N^{pair}_{data}  \rho^{II}_{m}(\Delta\eta,\Delta\phi)}
\end{eqnarray}

\begin{eqnarray}
where, \hspace{2ex} \rho^{II}_{d}(\Delta\eta,\Delta\phi) = \frac{d^{2}N_{data}}{d\Delta\eta d\Delta\phi} \hspace{2ex} and \hspace{2ex} \rho^{II}_{m} = \frac{d^{2}N_{mixed}}{d\Delta\eta d\Delta\phi}
\end{eqnarray}

\noindent are the distributions of pairs of particles from the data and mixed events respectively. $\rho^{II}(\Delta\eta,\Delta\phi)$ distributions for the data and mixed event were generated by counting the number of pairs in the interval $\Delta\eta$ and $\Delta\phi$. For evaluating C($\Delta\eta$,$\Delta\phi$), the distributions for the data/mixed events were normalized to the number of pairs ($N^{pair}_{data}$ or $N^{pair}_{mixed}$). The calculations are performed considering the charged particles of each event having $\phi$ values lying in the interval $0<\phi<2\pi$ and $\eta$ values lying in the range $|\eta| \leq \eta_{c} \pm 3.0$, $\eta_{c}$ being the centre of symmetry of the $\eta$ distribution.\\

\noindent The correlation function C($\Delta\eta$,$\Delta\phi$) is estimated for the five events selected on the basis of the criteria discussed in the previous section and are plotted in Figs.6 and 7; the different scales of Z-axis be noted for the plots for Evts$\#$ 5, 16, 21 and 22 shown in Fig.6 and that for Evts$\#$ 27 and 22, displayed in Fig.7. The correlation function for Evt$\#$ 22 (the 'cold' event) is plotted in both the figures on different Z-scales of 'hot' events for comparison sake. The difference is noticeable due to the different z-scales chosen in the two figures. The z-scales in the two figures are kept the same as that for the other events in the figure so that the fluctuations in the ¡®hot¡¯ and ¡®cold¡¯ events be clearly reflected. The following references may be drawn from Figs.6 and 7.\\

\begin{enumerate}
\item In the 'cold' event (Evt$\#$22), shown in the bottom panel of the figures, presence of two particle correlations is noticed in the region of $\Delta\eta$ ($\eta_{c} \pm 2$) and through out the $\phi$ region considered. This indicates the presence of two particle correlations in $^{208}$Pb-$^{208}$Pb collisions at 158A GeV/c.

\item Evt$\#$ 21 shows few distinct peaks around $\eta_{c}$ ($\sim$ 3.5) which spreads in considerable range of $\phi$. It is interesting to mention that the F$_{2}$ values of this event are significantly large in the $\eta$-space while in the $\phi$-space its F$_{2}$ values are not so large and compare well with the event averaged values (Fig.2).

\item Evts$\#$ 5 and 16 exhibit some distinct peaks at $\Delta\eta \sim 1.5$ and 3.5 and 6.0. the magnitude of correlation, C($\Delta\eta$,$\Delta\phi$) in these regions are as large as $\sim$ 6.

\item Evt$\#$ 27 at $\Delta\eta \sim 1.0$ gives very prominent peaks where the magnitude of correlations function are as high as $\sim$ 16.

\item The F$_{2}$ values of the Evts$\#$ 5, 16 and 27 which exhibit strong two-particle correlations may also be noted to be quite large in $\eta$, $\phi$ and $\eta$-$\phi$ spaces. These events have been noticed to show dominant peaks and valleys in their $\eta$ and $\phi$ distributions too.

\end{enumerate}

\noindent{\bf 3. Summary}\\
\noindent In the present article, we have studied the fluctuation observables relevant to investigation of $^{208}$Pb-$^{208}$Pb collisions at 158 A GeV/c. Intermittency or factorial moment analysis applied to individual high multiplicity events may be used to identify events with densely populated narrow phase space regions. The four events selected on the basis of their F$_{2}$ values are found to exhibit large fluctuations in $\eta$ and $\phi$ distributions and exhibit strong two-particle correlations. The magnitude of the correlation function in few of the events appear to be as high as $\sim$ 16. These observations give clear indication that if the system created during the collision has undergone phase transition, the associated fluctuation observables can be reliably analyzed on the basis of SFMs value. The result of the present analysis thus suggest that if the system formed during the collision has undergone the phase transition, the associated  events identified on the basis of their SFMs values can be analyzed individually in detail which may lead to arrive at draw some interesting conclusions.\\


\noindent{{\bf References}}
\begin{enumerate}
\item[1] Shakeel Ahmad et al., Int. J. Mod. Phys. E23 (2014) 1450065.
\item[2] S. A. Voloshin et al., Phys. Rev. C60, (1999) 024901.
\item[3] M. Doring and V. Koch, Acta Phys. Pol. B33 (2002) 1495.
\item[4] L. D. Landau and E. M. Lifshitz, Statistical Physics (Peragamon Press, 1958).
\item[5] E. Shuryak, J. Phys. G35 (2008) 104044.
\item[6] H. Hess et al., Phys. Rev C 73 (2006) 034913.
\item[7] K. Adcox et al., Phys. Rev C 66 (2002) 024901.
\item[8] Shakeel Ahmad et al., Phys. Scr. 87 (2013) 045201.
\item[9] S. S. Adler et al., (PHENIX Collab.), Phys. Rev. Lett. 93 (2004) 092301.
\item[10] D. Adamova et al., (CERES Collab.), Nucl. Phys. A727 (2003) 97.
\item[11] F. Jinghua and L. Lianshou, arXiv:0310308v1[hep-ph] (2003).
\item[12] L. Lianshou and F. Jinghua, arXiv:0401129v1[hep-ph] (2004).
\item[13] C. Alt et al., Phys. Rev. C78 (2008) 034914.
\item[14] S. Bhattacharyya et al., Phys. Rev. Lett. B726 (2013) 1350066.
\item[15] M. L. Cherry et al., (KLM Collab.), Acta Phys. Pol. B29 (1998) 2129.
\item[16] Shakeel Ahmad et al., Euro. Phys. Lett. 112 (2015) 42001.
\item[17] T. Anticic et al., (NA49 Collab.), Phys. Rev. C81 (2010) 064907.
\item[18] A. Bialas and B. Ziaja, Phys. Lett. B378 (1996) 319.
\item[19] S. Mrowczynski, Acta Phys. Pol B40 (2009) 1053.
\item[20] L. Lianshou, (EMU01 Collab.), Nucl. Phys. B (Proc. Suppl.) 71 (1999) 341.
\item[21] K. Fialkowski and R. Wit, Acta Phys. Pol. B31 (2000) 681.
\item[22] K. Fialkowski and R. Wit, Acta Phys. Pol. B30 (1999) 2759.
\item[23] M. Hauer et al., J. Phys. G. 35 (2008) 044064.
\item[24] D. Ghosh. et al., Chin. Phys. Lett. 19 (2002) 1436.
\item[25] S. L. C. Barroso et al., arXiv:astro-ph.HE/1011.3764v1.
\item[26] Y. F. Wu and L. S. Liu, Phys. Lett. B269 (1999) 28.
\item[27] L. S. Liu, (EMU01 Collab.), Nucl. Phys. B71 (1999) 341.
\item[28] M. I. Adamovich et al., (EMU01 Collab.), J. Phys. G22 (1996) 1469.
\item[29] M. I. Adamovich et al., (EMU01 Collab.), Phys. Lett. B217 (1989) 285.
\item[30] M. I. Adamovich et al., (EMU01 Collab.), Phys. Rev. Lett. 65 (1990) 412.
\item[31] M. I. Adamovich et al., (EMU01 Collab.), Phys. Lett. B201 (1988) 397.
\item[32] M. I. Adamovich et al., (EMU01 Collab.), Phys. Lett. B263 (1991) 539.
\item[33] A. Bialas and R. Peschanski, Nucl. Phys. B273 (1986) 703.
\item[34] I. V. Azhinenko et al., (NA22 Collab.), Phys. Lett. B222 (1989) 306. 
\item[35] R. Holynski et al., (KLM Collab.), Phys. Rev. Lett 62 (1989) 733.
\item[36] E. A. De. Wolf et al., Phys. Rep.270 (1996) 1.
\item[37] Shakeel Ahmad et al., J. Phys. G30 (2004) 1145.
\item[38] Shakeel Ahmad et al., Heavy Ion Phys. 25 (2006) 105.
\item[39] R. C. Hwa, Acta Phys. Pol B27 (1996) 531.
\item[40] B. Wosiek, Acta Phys. Pol. 46 (1996) 551.
\item[41] A. Bialas and R. Peschanski, Nucl. Phys. B308 (1988) 857.
\item[42] B. I. Abelev et al., Phys. Rev. Lett. 103 (2009) 172301.
\item[43] Shakeel Ahmad et al., Int. J. Mod. Phys. E22 (2013) 1350066.	
\item[44] B. Alver et al., (PHOBOS Collab.), Phys. Rev. C75 (2007) 054913.
\item[45] R. E. Ansorge et al., (UA5 Collab.), Z. Phys. C37 (1998) 191.
\item[46] A. Aduszkiewicz et al., arXiv:161000482v1[nucl-ex] (2016).
\item[47] K. Eggert et al., Nucl. Phys. B86 (1975) 201.
\item[48] G. Agakishiev et al., Phys. Rev. C86 (2012) 064902.
\end{enumerate}

\newpage
\begin{figure}[]
\begin{center}\mbox{\psfig{file=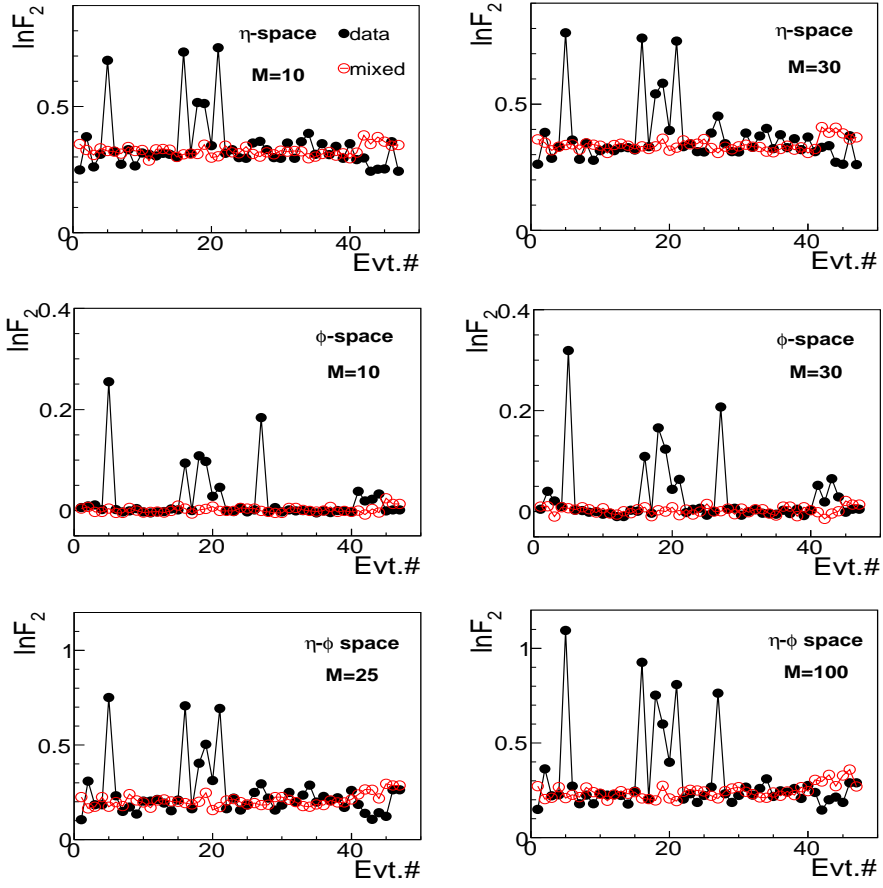,width=12cm,height=12cm}}
\end{center}
\caption[Fig1.]{\sf Event-wise variations of \(lnF_{2}\) in $\eta$-space (first row), $\phi$-space (middle row) and $\eta$-$\phi$ space (bottom row) for the data and mixed events produced in $^{208}$Pb-$^{208}$Pb collisions at 158A GeV/c.}
\label{lindat}
\end{figure} 

\newpage
\begin{figure}[]
\begin{center}\mbox{\psfig{file=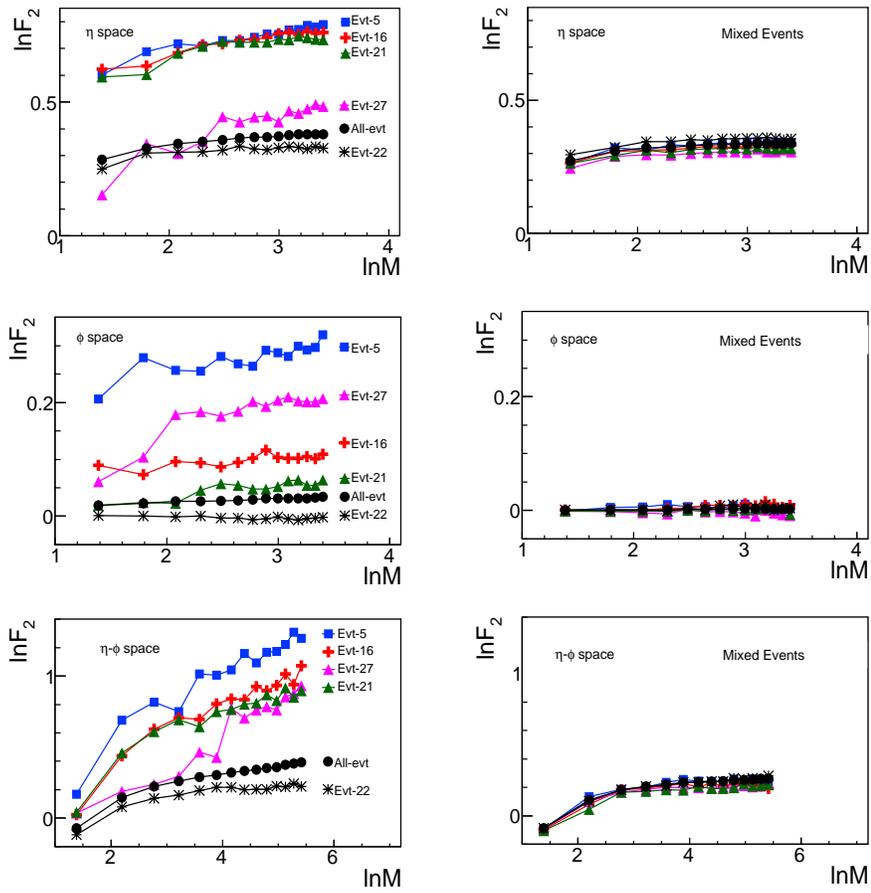,width=12cm,height=12cm}}
\end{center}
	\caption[Fig1.]{ {\sf Variations of \(lnF_{2}\) with \(lnM\) in $\eta$-, $\phi$- and $\eta$-$\phi$ spaces for all and five individual events produced in 158A GeV/c$^{208}$Pb-$^{208}$Pb interactions. Results from the mixed eventsare shown (right column)} }
\label{lindat}
\end{figure} 

\newpage
\begin{figure}[]
\begin{center}\mbox{\psfig{file=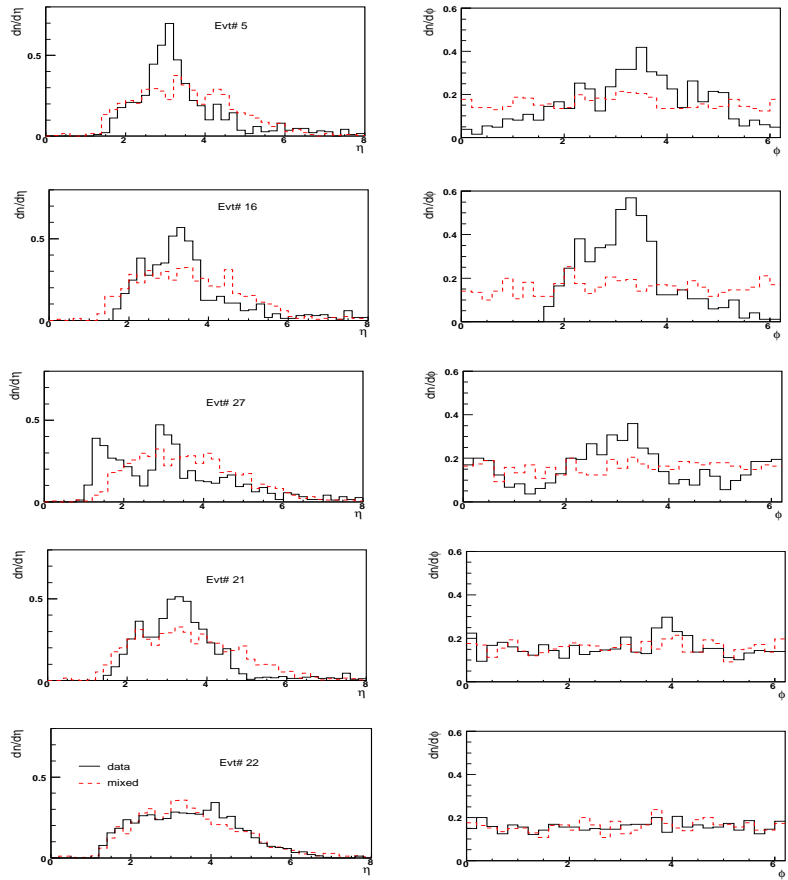,width=11cm,height=12cm}}
\end{center}
	\caption{\sf $\eta$ and $\phi$ distributions for the four "hot" and one "cold" 158A GeV/c$^{208}$Pb-$^{208}$Pb collision events (solid lines) and corresponding mixed events (broken lines). }
\label{lindat}
\end{figure}

\newpage
\begin{figure}[]
\begin{center}\mbox{\psfig{file=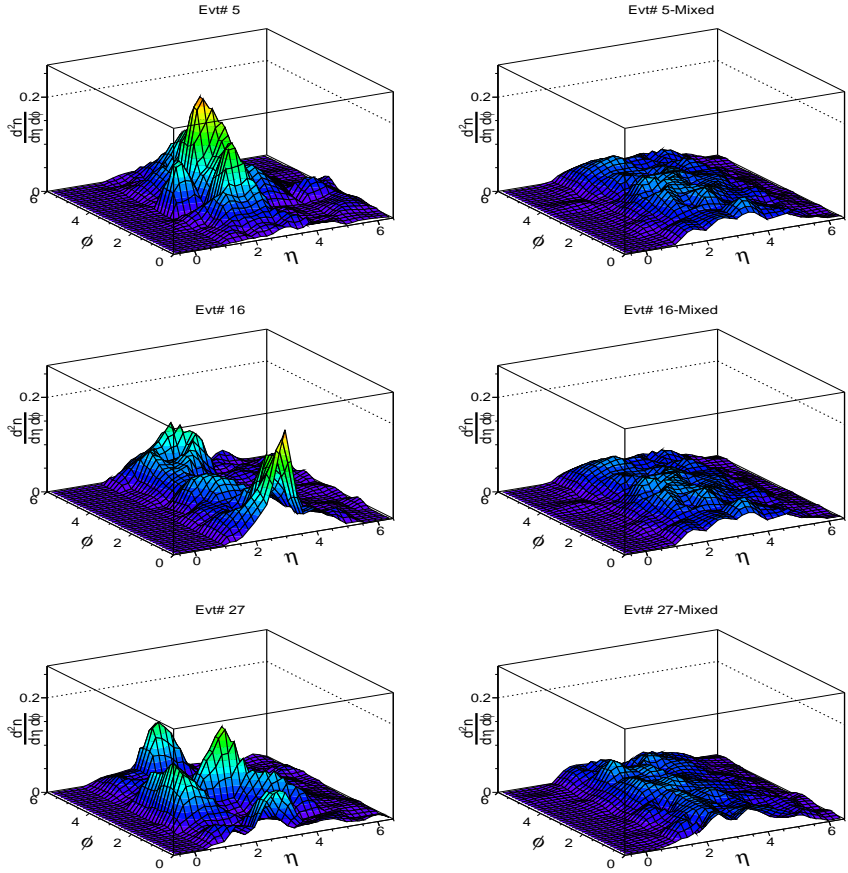,width=12cm,height=12cm}}
\end{center}
\caption{\sf Two-dimensional $\eta$-$\phi$ distributions  for Events $\#$ 5, 16 and 27 produced in 158A GeV/c $^{208}$Pb-$^{208}$Pb collisions (left column) and corresponding mixed events (right column).}
\label{lindat}
\end{figure}

\newpage
\begin{figure}[hptb]
\begin{center}\mbox{\psfig{file=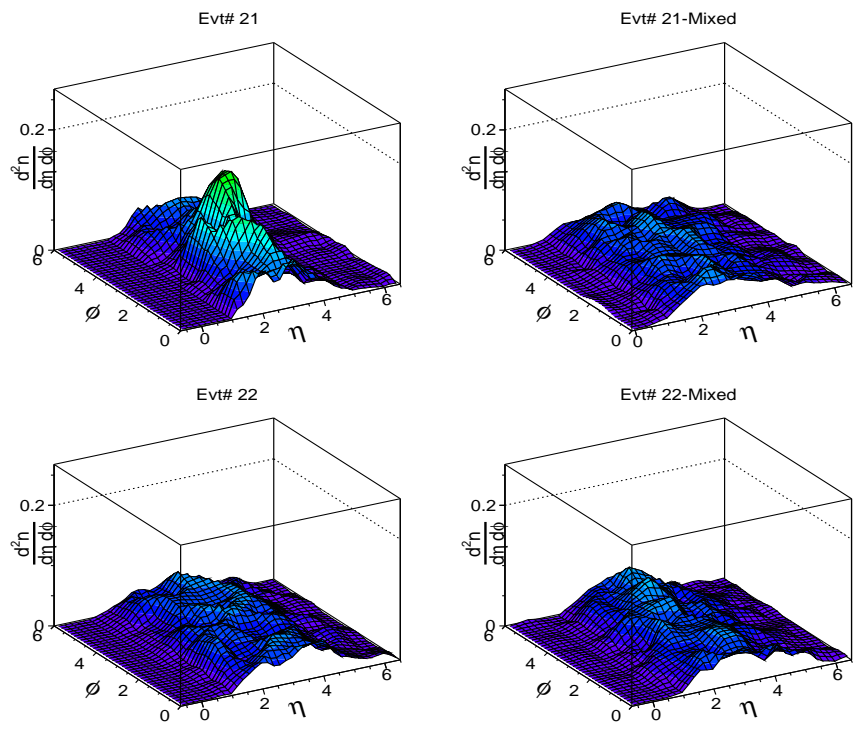,width=12cm,height=10cm}}
\end{center}
\caption{\sf The Same plots as in Fig.4 but for the Evts$\#$ 21 and 22.}
\label{lindat}
\end{figure} 

\newpage
\begin{figure}[]
\begin{center}\mbox{\psfig{file=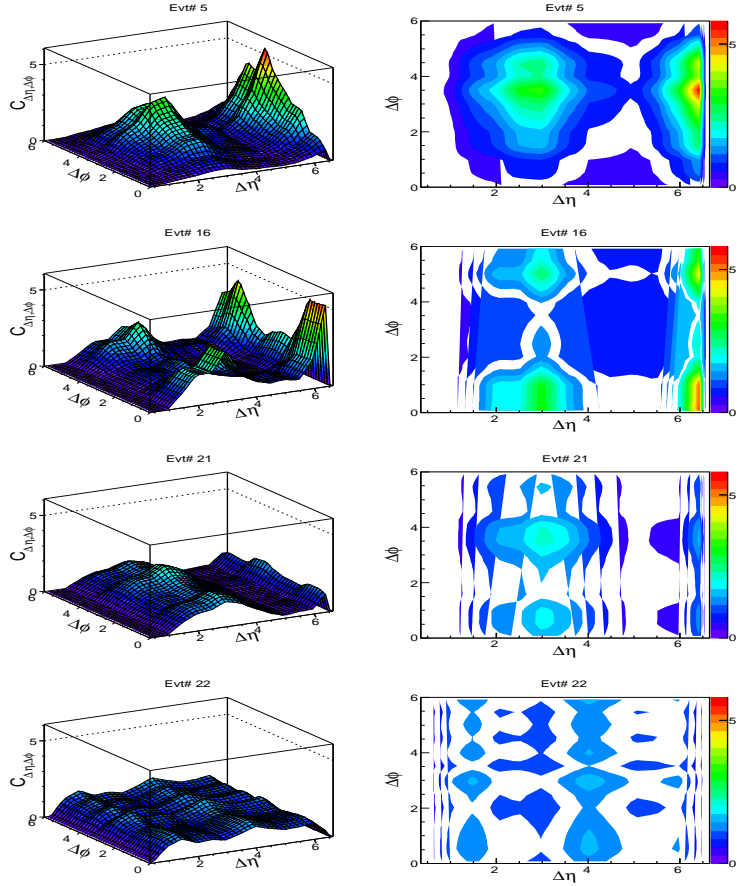,width=10cm,height=12cm}}
\end{center}
\caption{\sf Correlation function C($\Delta\eta$,$\Delta\phi$) for the Evts$\#$ 5, 16, 21 and 22 produced in $^{208}$Pb-$^{208}$Pb collisions at 158A GeV/c. The plots shown in the right column are the contour plots corresponding to those shown in the left column. }
\label{lindat} 
\end{figure}

\newpage
\begin{figure}[hptb]
\begin{center}\mbox{\psfig{file=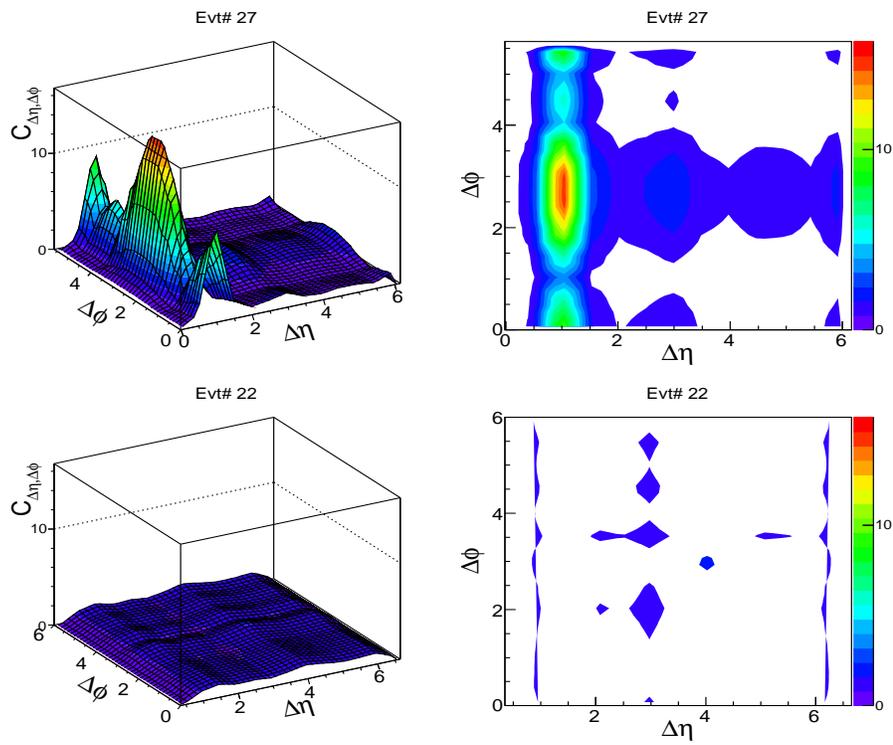,width=12cm,height=10cm}}
\end{center}
\caption{\sf Same plot as in Fig.6 but for Evts$\#$ 22 and 27: Results from Evt$\#$ 22, shown in Fig.6 are replotted here but on z-scale, different from that in Fig.6, for comparison with Evts$\#$ 27.}
\label{lindat}
\end{figure} 

\end{document}